\documentclass[a4paper]{article}

\usepackage{graphicx}
\usepackage{amsfonts}
\usepackage{amsmath}
\usepackage[a4paper]{geometry}
\usepackage{layouts}
\usepackage{float}

\title{Efficient coding of spectrotemporal binaural sounds leads to emergence of the auditory space representation}
\author{Wiktor M\l ynarski\\Max-Planck Institute for Mathematics in the Sciences\\mlynar@mis.mpg.de}
\date{}

\begin{document}
\maketitle

\begin{abstract}

To date a number of studies have shown that receptive field shapes of early sensory neurons can be reproduced by optimizing coding efficiency of natural stimulus ensembles. A still unresolved question is whether the efficient
coding hypothesis explains formation of neurons which explicitly represent environmental features of different functional importance. 
This paper proposes that the spatial selectivity of higher auditory neurons emerges as a direct consequence of learning efficient codes for natural binaural sounds. Firstly, it is demonstrated that a linear efficient coding transform - Independent Component Analysis (ICA) trained on spectrograms of naturalistic simulated binaural sounds extracts spatial information present in the signal. A simple hierarchical ICA extension allowing for decoding of sound position is proposed. Furthermore, it is shown that units revealing spatial selectivity can be learned from a binaural recording of a natural auditory scene. In both cases a relatively small subpopulation of learned spectrogram features suffices to perform accurate sound localization. Representation of the auditory space is therefore learned in a purely unsupervised way by maximizing the coding efficiency and without any task-specific constraints. 
This results imply that efficient coding is a useful strategy for learning structures which allow for making behaviorally vital inferences about the environment.
\end{abstract}

\section{Introduction}
As originally proposed by \cite{Barlow1}, the efficient coding hypothesis suggests that sensory systems adapt to the statistical structure of the natural environment in order to maximize the amount of conveyed information. This implies that stimulus patterns encoded by sensory neurons should reflect statistics and redundancies present in the natural stimuli. Indeed - it has been demonstrated that learning efficient codes of natural images \cite{Olshausen, BellSejnowski} or sounds \cite{Lewicki} reproduces shapes of neural receptive fields in the visual cortex and the auditory periphery. Additionally, recent studies provided statistical evidence suggesting that spectrotemporal receptive fields (STRFs) at processing stages beyond the cochlea, are adapted to the statistics of the auditory environment \cite{Carlson, Terashima, Klein} to provide its efficient representation.
However, having a sole representation of the stimulus is not enough for the organism to interact with the environment. In order to perform actions, the nervous system has to extract relevant information from the raw sensory data and then segregate it according to its functional meaning. For example the auditory system must extract position invariant information regardless of sound quality, separating ''what'' and ''where'' information.  In a more recent paper \cite{Barlow2001} Barlow proposed that behaviorally relevant stimulus features (i.e. ones supporting informed decisions) may be learned by redundancy reduction. In other words, functional segregation of neurons can be achieved by  efficient coding of sensory inputs. The evidence in support of this notion is still sparse.

Among different sensory mechanisms, spatial hearing provides a good example for the extraction and separation of behaviorally vital information from the sensory signal. The ability to localize and track sound sources in space is of a critical relevance for the survival of most animal species. Contrary to vision, audition covers the entire space surrounding the listener and may therefore provide an early warning about the presence or motion of objects in the environment. Mammals localize sounds on the azimuthal plane using two ears \cite{Grothe, MoreThanOneEar}. Binaural hearing mechanisms rely on between-ear disparities to infer the spatial position of the sound source. In humans, according to the well known Duplex Theory \cite{Strutt, Grothe}, interaural time differences (ITDs) constitute a major cue for low frequency sound localization and sounds of high frequency ($>1500$ Hz) are localized with interaural level differences (ILDs). However in natural hearing conditions, spectrotemporal properties of sounds vary continuously, hence combinations of cues available to the organism also change.
Even though temporal differences on the order of microseconds are of a substantial importance for sound localization, binaural neurons in the higher areas of the auditory pathway can be characterized with Spectrotemporal Receptive Fields (STRFS), which have much more coarse temporal resolution (ms) \cite{Gill}. Despite such loss of temporal accuracy, many of those neurons reveal sharp spatial selectivity \cite{Schnupp} encoding the position of the sound source in space. What is the neural computation underlying this process remains an open question.

The present paper uses spatial hearing as an example of a sensory task, to show how information of different meaning (''what'' and ''where'') can be clearly separated. As its main result, it provides computational evidence pointing that redundancy reduction leads to the separation of spatial information from the representation of the sound spectrogram. This means that formation of the neural auditory space representation can be achieved without the need of any task-specific computations but solely by applying the general principle of redundancy reduction.
It is demonstrated that Independent Component Analysis (ICA) - a linear efficient coding transform \cite{HyvBook} trained on a dataset of spectrograms of simulated as well as natural binaural speech sounds, extracts sound position invariant features separating them from the representation of the sound position itself. Learned structures can be understood as model spatial and spectrotemporal receptive fields of auditory neurons which encode different kinds of behaviorally relevant information. Current results are in line with known physiological phenomena and allow to make new experimental predictions.

\section{Material \& Methods}
High order statistics of natural auditory signal were studied by performing Independent Component Analysis (ICA) on a time-frequency representation of binaural sounds.

As a proxy for natural sounds, speech was used in the present study. Speech comprises a rich variety of acoustic structures and has been successfully used to learn statistical models predicting properties of the auditory system \cite{SmithLewicki, Carlson, Klein}. Additionally, it has been suggested that speech may have evolved to match existing neural representations, which are optimizing information transmission of environmental sounds \cite{SmithLewicki}.

Spatial sounds were obtained in two ways. Firstly, the efficient coding algorithm was trained using simulated naturalistic binaural sounds. Simulation gave the advantage of labeling each sound with its spatial position. Secondly a natural auditory scene was recorded with binaural microphones. The signal obtained in this way was less controlled, however it contained more complex and fully natural spatial information. Training datasets were obtained by drawing $70000$ random intervals $216$ ms long from each dataset separately.
The data generation process together with its interpretation is displayed on figure \ref{fig:01}.

\subsection{Simulated sounds}
As a corpus of natural sounds, data from the International Phonetic Association Handbook \cite{IPA} were used. The database contains speech sounds of a narrative told by male and female speakers in $29$ languages. All sounds were downsampled to $16000$ Hz from their original sampling rate and bandpass filtered between $200$ and $6000$ Hz. The training dataset was created by drawing random intervals of $216$ ms from the speech corpus data. 
Spatial sounds were simulated by convolving sampled speech chunks with human Head Related Transfer Functions (HRTFs). HRTF fully describe the sound distortion due to the filtering by the pinnae and therefore contain entire spatial information available to the organism. Given an angular sound source position $\theta$, HRTF is defined by a pair of linear filters:
\begin{equation}
 \label{hrtfDefn}
HRTF(\theta) = \{h_{L,\theta}(t), h_{R,\theta}(t)\}
\end{equation}
where $L,R$ subscrpits denote left and right ear respectively, and $t$ denotes time sample. One should note that in the temporal domain, HRTFs are often called Head Related Impulse Response (HRIR).
A set of HRTFs was taken from the LISTEN database \cite{Listen}. The database contains human HRTFs recorded for $187$ positions in the three-dimensional space surrounding the subject's head. HRTFs from a single random subject were selected and further limited to positions lying on the azimuthal plane with $15$ degree spacing ($24$ positions in total). Monaural stimulus vectors $x_E(t)$ ($E \in \{R,L\}$ denotes the ear) were created by drawing random chunks $g(t)$ of speech sounds and convolving them with $HRTF(\theta)$ corresponding to an azimuthal position $\theta$, which was also randomly drawn: 
\begin{equation}
\label{convolution}
x_E(t)=(g\ast h_E)(t)=\int_{-\infty}^{\infty} h_E(\tau)g(t-\tau)d\tau
\end{equation}
where $\ast$ denotes the convolution operator. In this data, spatial and identity information constitute independent factors.

\subsection{Natural sounds}
In order to obtain a dataset of natural binaural sounds a complex auditory scene was recorded using binaural microphones.
The recording consisted of three people (two males and one female) engaged in a conversation while moving freely in an echo-free chamber. Such an environment without reflections and echoes reduced the number of factors modifying sound waveforms. One of the male speakers was recording the audio signal with Soundman OKM-II binaural microphones placed in the ear channels.
In total $20$ minutes were recorded and included moving and stationary, often overlapping sound sources. 
To test the spatial sensitivity of learned features a recording with a single male speaker was performed. He walked around the head of the recording subject with a constant speed following a circular trajectory while reading a book out loud, twice in the clockwise and twice anti-clockwise direction. The length of the testing dataset was $54 s$.

\subsection{Simulated cochlear preprocessing}
\label{sec:dataRed}
Before reaching the auditory cortex, where spatial receptive fields (SRFs) were observed \cite{Schnupp}, sound waveforms undergo a substantial processing. Since the modeling focus of the present study was beyond the auditory periphery, the data were preprocessed to roughly emulate the cochlear filtering (see the scheme on fig \ref{fig:01}).

Short Time Fourier Transform (STFT) was performed on each sound interval included in the training dataset. Each chunk was divided into $25$ overlapping windows each $16$ ms long. STFT spanned $256$ frequency channels logarithmically spaced between $200$ and $4000$ Hz (decomposition into arbitrary, non-linearly spaced frequency channels was computed using the Goertzel algorithm). Logarithmic frequency spacing was observed in the mammalian cochlea and seems to be a robust property across species \cite{Greenwood, SmithChimeras}. The spectral power of the resulting spectrograms was transformed with a logarithmic function which emulates the cochlear compressive nonlinearity \cite{RoblesRuggero}. 

Stimuli were $216$ ms long in order to match the temporal extent of cortical neurons' STRFs, which were characterized by spatial receptive fields \cite{Schnupp}. Besides emulating the cochlear transformation of the air pressure waveform, such spectrograms  were reminiscent of the sound representation most effective in mapping spectrotemporal receptive fields in the songbird midbrain \cite{Gill}. A very similar representation was used in a recent sparse coding study \cite{Carlson}.  

Spectrograms of left and right ears were concatenated. Such data representation attempts to simulate the input to higher binaural neurons, which operate on spectrotemporal information, simulteneously fed from monaural channels \cite{Schnupp, Qiu, Miller}.  In principle, we could first train ICA on monaural spectrograms and then model their codependencies. In such way, however, the algorithm could not explicitly model binaural correlations. Additionally, this would require application of a hierarchical model, which lies outside of the scope of this study. Our approach resembles ICA studies, which focused on modelling of visual binocular receptive fields \cite{HoyerStereo, Hunt}. There, the input to binocular neurons in the visual cortex was modelled by concatenating image patches from the left and the right eye.

The efficient coding algorithm was run on the resulting time-frequency representation of the binaural waveforms.
After preprocessing the dimensionality of data vectors was equal to $2 \times (25 \times 256) = 12800$. Both training datasets: simulated and natural one consisted of $70000$ samples. Prior to the ICA learning, the data dimensionality was reduced with Principal Component Analysis (PCA) to $324$ dimensions, preserving more than $99\%$ of total variance in both cases. Due to memory issues (allocation of a very large covariance matrix) a probabilistic PCA implementation was used \cite{Roweis}.

\subsection{Independent Component Analysis}
Independent Component Analysis (ICA) is a family of algorithms which attempt to find a maximally non-redundant, information-preserving representation of the training data within the limits of the linear transform \cite{HyvBook}. In its standard version, given the data matrix $X \in \mathbb{R}^{n \times m}$ (where $n$ is the number of data dimensions and $m$ the number of samples), ICA learns a filter matrix $W \in \mathbb{R}^{n \times n}$, such
that:
\begin{equation}
WX=S
\end{equation}
where columns of $X$ are data vectors $\mathbf{x} \in \mathbb{R}^n$, rows of $W$ are linear filters $\mathbf{w} \in \mathbb{R}^n$ and $S \in \mathbb{R}^{n\times m}$ is a matrix of latent coefficients. 
Equivalently the model can be defined using a basis function matrix $A=W^{-1}$, such that:
\begin{equation}
\label{bfMatrix}
X=AS
\end{equation}
Columns $\mathbf{a} \in \mathbb{R}^n$ of matrix $A$ are known as basis functions and in neural systems modeling can be interpreted as linear stimulus features represented by neurons, which form an efficient code of the training data ensemble \cite{HyvBook}. Linear coefficients $s$ are in turn a statistical analogy of the neuronal activity. Since they can take both positive and negative values, their direct interpretation as physiological quantities such as firing rates is not trivial. For a discussion of relationship between linear coefficients and neuronal activity, please refer to \cite{HyvBook, RehnSommer}.
Equation \ref{bfMatrix} implies that each data vector can be represented as a linear combination of basis functions $\mathbf{a}$
\begin{equation}
\label{ICAsum}
x(t)=\sum_i s_i a_i(t)
\end{equation}
where $t$ indexes the data dimensions. The set of basis functions $\mathbf{a}$ is called a dictionary.
ICA attempts to learn a maximally non-redundant code. For this reason latent coefficients $s$ are assumed to be statistically independent i.e.
\begin{equation}
\displaystyle
p(s)=\prod_{i=1}^n p(s_i)
\end{equation}
The marginal probability distributions $p(s_i)$ are usually assumed to be sparse (i.e. of high kurtosis), since natural sounds and images have intrinsically sparse structure \cite{OlshausenV1} and can be represented as a combination of a small number of primitives. In the present work logistic distribution of the form:
\begin{equation}
p(s_i|\mu,\xi)=\frac{\exp(-\frac{s_i-\mu}{\xi})}{\xi(1+\exp(-\frac{s_i-\mu}{\xi}))^2}
\end{equation}
with position $\mu=0$ and scale parameter $\xi=1$ is assumed. 
The basis functions are learned by maximizing the log-likelihood of the model via gradient ascent \cite{HyvBook}.

\subsection{Analysis of learned basis functions}
Similarity between left and right ear parts of learned basis functions was assessed using the Binaural Similarity Index (BSI), as proposed in \cite{Miller}. The BSI is simply Pearson's correlation coefficient between left and right ear parts of each basis function. BSI equal to $-1$ means that absolute values at every frequency and time position are equal and have the opposite sign, while BSI equal to $1$ means that the basis function represents the same information in both ears

Dictionary of binaural basis functions learned from natural data was classified according to the modulation spectra of their left ear parts. A modulation spectrum is a two-dimensional Fourier transform of a spectrogram. It is informative about spectral and temporal modulation of learned features and it has been applied to study properties of natural sounds \cite{TheunissenMod} and real \cite{Miller} as well as modeled \cite{Saxe} receptive fields in the auditory system.

Spatial sensitivity of basis functions learned from natural data was further quantified by means of Fisher information. Fisher information is a measure of how accurate one can estimate a hidden parameter $\theta$ from an observable $s$ knowing a conditional probability distribution $p(s|\theta)$ \cite{BrunelNadal}. Here, $\theta$ corresponds to the angular position of the auditory stimulus and $s$ to one of the sparse coefficients. Assuming a deterministic mapping $s(\theta)=f(\theta)=\mu_\theta$ distorted with a zero-mean stationary Gaussian noise, one obtains:
\begin{equation}
\displaystyle
p(s|\theta)=\mathcal{N}(s|\mu_\theta, \sigma)
\end{equation}.
For simplicity $\sigma$ was assumed to be equal to $1$. Fisher information $\mathcal{I}(\theta)$ then becomes \cite{BrunelNadal}:
\begin{equation}
\label{fisherInfo}
\displaystyle
\mathcal{I}(\theta) = (\frac{d}{d\theta}f(\theta))^2
\end{equation}
Mean values $\mu_\theta$ were estimated by averaging coefficient activations over four trials during which the speaker walked around the head of the subject. Each activation time course was additionally smoothed with a $20$ samples long rectangular window. 

\section{Results}
Besides the properties of the sound source itself, natural sounds reaching the ear membrane are also shaped by head-related filtering. The spectrotemporal structure imposed by the filter depends on the spatial configuration of objects. By performing redundancy reduction the auditory system could, in principle, separate those two sources of variability in the data and extract spatial information. 
One should observe that transformations performed by the cochlea can strongly facilitate this task. The stimulus $x_E$ (where $E \in {L,R}$ indicates the left or the right ear) is an air pressure waveform $g(t)$ convoluted with an HRTF (or a combination of HRTFs) $h_{E,\theta}(t)$, as defined by equation \ref{convolution}.  The basilar membrane performs frequency decomposition, emulated here by the Fourier transform:
\begin{equation}
\label{fourier}
\displaystyle
\mathcal{F}(x,\omega)=\int_{-\infty}^{\infty}x_E(t)\exp(-2\pi i \omega t) dt = A^x_\omega(\cos \phi^x_{E,\omega} + i \sin \phi^x_{E,\omega})
\end{equation}
where $\omega$ denotes frequency, $A^x_{E,\omega}$ amplitude and $\phi^x_{E,\omega}$ phase. By the convolution theorem \cite{Katznelson}, convolution in the temporal domain is equivalent to a pointwise product in the frequency domain, i.e.
\begin{alignat*}{2}
\label{convTheorem}
\mathcal{F}((g\ast h_E),\omega) &= \int_{-\infty}^{\infty}g(t)\exp(-2\pi i \omega t) dt \int_{-\infty}^{\infty}h_E(t)\exp(-2\pi i \omega t) dt =\\ &= A^g_{\omega}(\cos \phi^g_\omega + i \sin \phi^g_\omega) A^h_{E,\omega}(\cos \phi^h_{E,\omega} + i \sin \phi^h_{E,\omega})
\end{alignat*}
Additionally, the basilar membrane applies a compressive nonlinearity \cite{RoblesRuggero} which this study approximates by transforming the spectral power with a logarithmic function. Since the logarithm of the product is equal to the sum of logarithms, the spectral amplitude of the stimulus $A^x_{E,\omega}=A^h_{E,\omega} A^g_\omega$ can be decomposed into the sum:
\begin{equation}
\label{logsum}
\log(A^h_{E,\omega} A^g_\omega)=\log(A^h_{E,\omega}) + \log(A^g_\omega) 
\end{equation}.
This means that the spectrotemporal representation of the signal generated by the cochlea is a sum of the raw sound and HRTF features. One should note, however, that the above analysis applies to an infnite window Fourier transform, and the data used in this study was generated by performing a Short Time Fourier Transform (STFT) with a $16$ ms long, overlapping windows. Fourier coefficients were mixed between neighboring windows due to their overlap. For point-source, stationary sounds this effect did not influence the $\log(A^h_{E,\omega})$ term of the equation \ref{logsum}, since HRTFs were shorter than the STFT window, hence hear-related filtering was temporally constant. For a dynamic scene, where neighboring STFT windows contained different spatial information, the additive separability of sound and HRTF features (as described by equation \ref{logsum}) may have been distorted. Taken together, a linear redundancy reducing transform such as ICA provides a reasonable approach to separate information about object positions from the raw sound. In an ideal case, ICA trained on stimulus spectrograms $A^x_\omega$ could separate representation of HRTF ($A^h_{E,\omega}$) and stimulus ($A^g_{E,\omega}$) amplitudes into two distinct basis functions sets \cite{HarperOlshausen}. The difficulty of the separation task depends on the temporal variability of the spatial information which reflects configuration of the environment (i.e. number of sources, their motion patterns and positions).
The current study considers two cases of different complexity: (a) simulated dataset consisting of short periods of speech displayed from single positions and (b) a binaural recording of a natural scene with freely moving human speakers.

\subsection{Simulated sounds}
The goal of the present study was to identify high-order statistics of natural sounds informative about positions of the sound source. Association of a sound waveform with its spatial position requires detailed knowledge about source localization i.e. each sound should be labeled with spatial coordinates of its source. For this reason binaural sounds studied in this section were simulated, using speech sounds and human HRTFs. Naturalistic data created in this way resembled binaural input from the natural environment, while making position labeling of sources available.

From the simulated dataset, after reducing data dimensionality with PCA (see section \ref{sec:dataRed}), $324$ ICA basis functions were learned. A subset of $100$ features is depicted in fig \ref{fig:02}. It is clearly visible that the learned basis can be divided into two separate subpopulations by the similarity between their left and right ear parts, which is quantified by the Binaural Similarity Index (BSI) (see Materials and Methods). Sorted values of the BSI are displayed on fig \ref{fig:06}A as black circles. The majority of basis functions ($314$) exceed the $0.9$ threshold and only $10$ fall below it. Out of those $8$ reveal strong negative interaural correlation and only $2$ are close to $0$. Basis functions  with the BSI below $0.9$, were separated from the rest and all ten of them are depicted on fig \ref{fig:02}A. Since they represent different information in each ear they are going to be called ''binaural'' through the rest of the paper. This is in contrast to ''monaural'' basis functions which encode similar sound features in both ears (see fig \ref{fig:02}(B))

The binaural sub-dictionary captures signal variability present due to the head-related filtering. Even though the training dataset included sounds displayed from $24$ positions, hence $24$ different HRTFs were used, only $10$ binaural basis functions emerged from the ICA. Out of those, almost all are temporally stable i.e. do not reveal any temporal modulation (except for $2$ - positions $5$ and $6$ on fig \ref{fig:02} A). The dominance of temporally constant features was expected, since training sounds were displayed from fixed positions and were convoluted with filters, which did not change in time. Temporally stable basis functions weight spectral power across frequency channels, mostly with opposite sign in both ears (as reflected by negative values of the BSI). Surprisingly, despite the lack of moving sounds in the training dataset, two temporally modulated basis functions were also learned by the model. They represent envelope comodulation in high frequencies with an interaural phase shift of $\pi$ radians.

A representative subset of $90$ monaural basis functions is depicted on fig \ref{fig:02}B. Their left and right ear parts are exactly the same and encode a variety of speech features. Regularities such as harmonic stacks, on- and offsets or formants are visible. Captured monaural patterns essentially reproduce results from a recent study by \cite{Carlson} which shows that efficient coding of speech spectrograms learns features similar to STRFs in the Inferior Colliculus. Monaural basis functions are, however, not a focus of the present study and are not going to be discussed in detail. 

A separation of the learned dictionary into two subpopulations of binaural and monaural basis functions ($a^g$ and $a^k$ respectively) allows to represent every sound spectrogram in the training dataset as a linear combination of two isolated factors i.e. representations of speech and HRTF structures (see fig \ref{fig:02} (C)). Taking this fact into account, equation \ref{ICAsum} can be rewritten as:
\begin{equation}
\label{ICAdecomp}
x(t)=\sum_{i=1}^G s^g_i a^g_i(t) + \sum_{j=1}^K s^k_j a^k_j(t)
\end{equation}
This notation explicitly decomposes the basis into $G$ spatial basis functions $a^g$ and $K$ and non-spatial basis functions $a^k$.

\subsubsection{Emergence of model spatial receptive fields}
Marginal coefficient histograms conformed rather well to the logistic distribution assumed by the ICA model, although binaural coefficients were typically more sparse (see figure \ref{fig:03}). In order to understand how informative learned features are about position of sound sources, conditional distributions of the linear coefficients were studied. Histograms conditioned on a location of a sound source reveal whether any spatial information is encoded by learned basis functions.

Fig \ref{fig:03} (A)-(F) displays $6$ basis functions and corresponding conditional histograms. The horizontal axis of each conditional histogram corresponds to the angular position of the sound source (from $0$ to $345$ degrees). A vertical cross-section is a normalized histogram of the coefficient values for all sounds displayed in the training dataset from a particular position (around $2900$ samples on average).

Three representative monaural basis functions are depicted on fig \ref{fig:03} (D)-(F). It is immediately visible that conditional distributions of their coefficients are stationary across spatial positions. The zero-centered logistic pdf with a constant scale parameter (parameters equal to those of the marginal pdf) is preserved across all positions.  This implies that coefficients of monaural basis functions are independent from the sound source location. Monaural bases encode speech features and since all speech structures were displayed from all positions in the training data, their activations do not carry spatial information. This property is characteristic for all basis functions with BSI greater than $0.9$.

Coefficients of binaural basis functions reveal a very different dependency structure (see fig \ref{fig:03} (A)-(C)). Their variance at each spatial position is very low, however, variability across positions is much higher. Activations of binaural features remain close to zero at most angular positions regardless of the sound identity. At few preferred positions they reveal pronounced peaks in activation (positive or negative) reflected by strong shifts in the mean value.   
This highly non-stationary structure of conditional pdfs is informative about the sound position, while remains almost invariant to the sound's identity (which is reflected by the small standard deviation). 
Basis function depicted on fig \ref{fig:03} A responds with a strong positive activation to sounds originating at $270$ degrees (i.e. directly in front of the right ear) and with a strong negative activation to sounds originating from the directly opposite location - at $90$ degrees (i.e. in front of the left ear). Sounds at positions deviating $+/- 15$ degrees from peaks also modulate basis activations, although activations are weaker. Similar spatial selectivity pattern is revealed by the basis function on fig \ref{fig:03} C, which however responds positively to sounds at $60$ and negatively to sounds at $315$ degrees.
The spectrotemporal feature on fig \ref{fig:03} B encodes spatial information of a particularly high behavioral relevance. Its activity significantly deviates from zero, only when sounds are placed behind the head in the interval between $165$ and $210$ degrees. This region is not visually accessible, therefore position or motion of objects in that area has to be inferred basing on auditory information only.
It may appear that conditional histograms are symmetric around the $180$ degree point. However, positive and negative peaks of coefficient histograms do not have exactly equal absolute values. 

It is important to notice here that each spectrotemporal feature captured by binaural basis functions is an indirect representation of the sound position in the surrounding environment. Therefore if ICA basis functions can be interpreted as STRFs of binaural neurons, the corresponding conditional histograms constitute a theoretical analogy of their spatial receptive fields (SRFs) informing the organism about the position of the sound source within the head-centered frame of reference.  

\subsubsection{Decoding of the sound position}
As described in the previous subsection, linear coefficients of binaural basis functions are informative about the location of the sound source. Spatial selectivity of single basis functions is however not specific enough to reliably localize sounds.
Pairwise coefficient activations of two exemplary basis functions are depicted on fig \ref{fig:03} G. Each point represents a single sound and its color corresponds to the source's angular position. Strong clustering of same-colored points is strongly visible. They form at least $6$ highly separable clusters. This, in turn, shows that the joint distribution of those two coefficients contains more information about the source position than one dimensional conditional pdfs. 
This is in contrast to fig \ref{fig:03} H depicting co-activations of two monaural basis functions. There, points of all colors are strongly mixed, creating a ''salt and pepper'' pattern, where no clear separation between source positions is visible. 

To test, whether reliable decoding of sound position from activations of binaural basis functions is possible, this work employs the Gaussian Mixture Model (GMM). The GMM models the marginal distribution of latent coefficients $s^g$ used for the position decoding as a linear combination of Gaussian distributions, such that:
\begin{equation}
\label{GMM1}
\displaystyle
p(s^g)=\sum_{k=1}^{24} p(s^g|C_k)p(C_k)
\end{equation}

\begin{equation}
\label{GMM2}
\displaystyle
p(s^g|C_k)=\mathcal{N}(s^g| \mu_k, D_k)
\end{equation}\\
where $C_k$ is a position label ($C_1=0 \deg, C_{24}=345 \deg$) and $\mu_k, D_k$ denote a position specific mean vector and covariance matrix respectively. The structure of dependencies among random variables is presented in a graphical form in fig \ref{fig:04} (A). Since the prior on position labels $p(C_k)$ is assumed to be uniform, the decoding procedure can be recast as a maximum-likelihood estimation:
\begin{equation}
\label{GMM3}
\displaystyle
\hat{C} = \arg\max_k p(s^g|C_k)
\end{equation}
where $\hat{C}$ is the decoded position. The resulting procedure iterates over all position labels and returns the one which maximizes the probability of an observed data sample. 

The decoding performance relies on the selected subset of basis functions used for this task. To test whether binaural features contribute stronger to the position decoding than monaural ones, all basis functions were sorted according to their BSI. Then, the GMM was trained using incrementally larger number of latent coefficients, starting from a single one corresponding to the basis function with the highly negative BSI and ending using the entire basis function set. In every step, for the GMM training $70\%$ of the data were used, while remaining $30\%$ were used for cross-validation.
The average decoder performance is plotted against the number of used features on figure \ref{fig:04} B. Binaural features are separated from the monaural ones with a dashed vertical line. A straightforward observation is that binaural basis functions almost saturate the decoding accuracy. Indeed it reaches the level of $97.9\%$. Adding remaining $314$ monaural basis functions increases the performance to $99.7\%$ which is only $1.8$ percentage point. Interestingly, temporally modulated binaural basis functions number $5$ and $6$ did not contribute to the decoding quality, which is visible as a short plateau on the plot.
Saturation of the decoder's performance by binaural basis function activations entails that almost entire spatial information present in the sound is separated from other kinds of information by the ICA model and represented by binaural basis functions. Relating this observation to the nervous system, this means, that the spatial position of natural sound sources can be decoded from the joint activity of a relatively small subpopulation of binaural neurons. 

\subsection{Natural sounds}
The previous section described results for simulated sounds. While simulated sounds have the advantage of giving a full control over source positions they are only a very crude approximation to the binaural stimuli occurring in the real natural environment. This section describes results obtained using binaural recordings of a real-world auditory scene, consisting of three speakers moving freely in an echo-free environment.

Binaurality of learned basis functions was again quantified with the BSI. Sorted BSI values are plotted on fig \ref{fig:06} A as gray triangles. A strong difference is visible, when compared with values of the dictionary trained on simulated data (black circles). Firstly, $64$ natural basis functions lay below the $0.9$ threshold - many more compared to only $10$ simulated ones. Secondly, natural BSIs vary more smoothly, and are more uniformly distributed between $-1$ and $0.9$ (see the histogram displayed in the inset). 

Similarly to the previous case, the learned dictionary was divided into two sub-dictionaries - binaural ones - below and monaural ones - above the $0.9$ BSI threshold. The sub-dictionary consisting of binaural basis functions is displayed on fig \ref{fig:05} A and fig \ref{fig:05} B displays $40$ exemplary monaural basis functions. While no qualitative difference is visible between monaural features when compared with results from the previous section (fig \ref{fig:02} B), the binaural sub-dictionaries differ strongly. Basis functions trained using natural data, reveal much richer variety of shapes including temporally modulated ones along patterns of strong spectral modulation.

\subsubsection{Properties of the learned representation}

This subsection presents properties of binaural basis functions trained with the natural binaural data. They were studied in more detail than the dictionary learned from simulated data since its structure is more complex and may reflect better the properties of binaural neurons. One should note that in neural systems modelling, neural receptive fields correspond better to ICA filters (rows $\mathbf{w}$ of matrix $W$ in equation \ref{ICAdecomp}). Basis functions, however, constitute optimal stimuli i.e. given basis function $\mathbf{a}_i$ as input the only non-zero coefficient is going to be $s_i$. Additionally, basis functions are a low-passed version of filters \cite{HyvBook}, and are more appropriate for plotting, since they represent actual parts of stimulus. For those reasons, this study focuses on basis function statistics.

The binaural dissimilarity of learned features was assessed with two measures. The BSI provides a continuous value quantifying how well the left ear part matches the right ear part. It however does not take into account the dominance of one ear over another. The dominance can be measured by comparing monaural peaks i.e. points of the maximal absolute value of left and right ear parts. Both measures were used by Miller and colleagues \cite{Miller} to describe receptive fields of binaural neurons in the auditory thalamus and cortex.
Monaural peaks (measured in standard deviation of the basis function dimensions) are compared on fig \ref{fig:06} (B). Crosses mark basis functions with the positive and diamonds with the negative BSI. Symbol sizes correspond to the absolute BSI value. 
Basis functions cluster along the diagonals (marked with dashed lines) which means that left and right ear peaks have similar absolute values and no clear dominance of a single ear is present. Interestingly, while roughly the same number of basis functions lays in upper right and both lower quadrants, only $4$ lay in the upper left one, corresponding to basis functions with a negative peak in the left ear and positive in the right ear.
Unfortunately, direct comparison of the analysis on fig \ref{fig:06} (B) with figure 9 in \cite{Miller} is not possible, due to the arbitrariness of the sign in the ICA model (coefficients can have positive and negative values, flipping the sign of the basis function). Additionally the notion of ipsi- and contra- laterality is meaningless for ICA basis functions. 

Shapes of basis functions belonging to the binaural sub-dictionary were studied by analyzing modulation spectra of their left-ear parts. Even though functions were binaural, classification according to only the single ear part was sufficient to identify subgroups with interesting binaural properties. Centers of mass of modulation spectra (for computation details see Materials and Methods) are plotted as circles on fig \ref{fig:06} (C). Circle color corresponds to the BSI value. Left parts of binaural features display a tradeoff between spectral and temporal modulation. This complies with the general trend of natural sound statistics \cite{TheunissenMod}.
Dictionary elements were divided into three distinctive groups according to their modulation properties (marked with roman numerals I, II, III and separated with dotted lines on fig \ref{fig:06} (C)). The first group consisted of weakly modulated features with spectral modulation below $0.3$ cycles/octave and temporal modulation below $4$ Hz. Majority of basis functions belonging to this group had high BSI, close to $0.9$. Three representative members of the first group are displayed on fig \ref{fig:06} (D) in the first row. Since their spectrotemporal modulation is weak, they capture constant patterns, similar in both ears, up to the sign.
The second group consists of basis functions revealing strong spectral modulation - above $0.3$ cycles/octave. Three exemplary members are visible in the second row of fig \ref{fig:06} (D). Basis functions belonging to the second group resemble majority of ones learned from simulated data. They weight spectral power across frequency channels constantly over time. In contrast to simulated basis functions, their BSIs are mostly close to $0$, indicating that channel weights do not necessarily have opposite sign between ears. Additionally, as visible in two out of three displayed examples, low frequencies below $1kHz$ are also weighted.

The third group includes highly temporally modulated features. Their temporal modulation exceeds $4$ Hz, while the spectral one stays below $0.3$ cycles/octave. Out of $15$ members of this group, only one has a positive BSI value - the rest remains close to $-1$. This implies that when their monaural parts are aligned with each other - corresponding dimensions have a similar absolute value and an opposite sign. Three examplary members of the third group are depicted in the last row of fig \ref{fig:06} (D). They are qualitatively similar to two temporal basis functions learned from the simulated data (they represent an envelope comodulation across multiple frequency channels with a $\pi$ phase difference). 

The temporal differences between monaural parts of basis functions were further studied using cross-correlation functions (ccf). Maximal values of the normalized ccf are plotted against maximizing temporal shifts on fig \ref{fig:06} (E). As in the fig \ref{fig:06} (C) - the color of circles represents the BSI value. The histogram of temporal shifts is depicted on fig \ref{fig:06} (F).
Cross-correlation of $30$ binaural features with a positive BSI, is maximized at $0$ temporal shift. In this case, BSI and the peak of cross-correlation have the same value. This is a property of basis functions with a weak temporal modulation, which constitute a major part of the binaural sub-dictionary. Features revealing temporal modulation have a negative BSI value (dark colors) and a non-zero temporal difference, which spanned the range between $-0.2$ to $0.2$ seconds.

\subsubsection{Spatial sensitivity of binaural basis functions}
In contrast to the simulated dataset, binaural recordings were not labeled with sound source positions. Furthermore, learned features may represent dynamic aspects of the object motion, therefore conditional histograms (constructed as in the previous section) would not be meaningful.

In order to verify whether binaural basis functions reveal tuning to spatial position of sound sources and invariance to their identity, a test recording was performed. One of the male speakers read a book out loud, while walking around the head of the recording subject, following a circular trajectory in a constant pace. This was repeated twice in the anticlockwise and twice in the clockwise direction. In such a way, the angular position of the speaker was made easy to estimate at each time point. The recording was divided into $216$ ms overlapping intervals, and each interval was encoded using the learned dictionary.
A general trend in the spatial sensitivity of basis functions was measured by computing correlation between estimated speaker's position and time courses of linear coefficients in the following way. Firstly, activation time courses were standarized to have mean equal to $0$ and variance equal to $1$. In the next step, time intervals where the coefficient's absolute value exceeded $1$ were extracted. This was done, since highly sensitive coefficients remained close to $0$ most of the time, and correlated with the speaker's position only in a narrow part of the space (i.e. their receptive field).
Elements of the binaural sub-dictionary correlated stronger with the estimated position than elements of the monaural one. Normalized histograms of linear correlations between the position of the sound source and sparse coefficients are presented on fig \ref{fig:07}. Monaural basis functions correlate much weaker with the sound position, which is reflected in the strong histogram peak around $0$. Binaural coefficients in turn, reveal strong correlations of the absolute value of $0.8$ in extreme cases. Linear correlation is however not a perfect way to assess relationship between sparse coefficients and the source position, since spatial selectivity of basis function may be limited to a narrow spatial area (as in fig \ref{fig:08} A and B). This results in correlations of low absolute values, even though spatial sensitivity of a basis function may be quite high.
To show spatial selectivity of learned features, their activations were plotted.
Resulting time courses of basis function activations are displayed as black continuous lines on fig \ref{fig:08}. Gray dashed lines mark approximated angular position of the speaker at every time point.

Subfigures (F)-(J) display activations of $5$ representative monaural basis functions. As expected, their activity correlates very weakly with the speaker's trajectory. Monaural basis functions encode features of speech and are invariant to the position of the speaker.
In contrary, activations of binaural basis functions visible on subfigures (A)-(E), reveal strong dependence on subjects position and direction of motion. Basis function A remains non-activated for most of positions and deviates from zero when the speaker is crossing the area behind the head of the recording subject. The slope of activation time courses is informative about the direction of speaker's motion. Similar, however noisier, spatial tuning is revealed by the basis function D. Basis function B displays broader spatial sensitivity, and its activation varies smoothly along the circle surrounding the subject's head. Spatial information represented by the spectrally modulated basis functions C and E does not have such a clear interpretation, however they display pronounced covariation with sound source's position (feature C for instance is strongly positively activated, when the speaker crosses directly opposite to the left ear).

Spatial sensitivity of basis functions can be further quantified using Fisher information (for computation details please see Materials and Methods). Figure \ref{fig:09} shows Fisher information estimates as a function of spatial position for features displayed on figure \ref{fig:08}. Each binaural basis function reveals a preferred region in space where source's position is encoded with higher accuracy. For this reason, histograms depicted on figures \ref{fig:09}A-E can be interpreted as an abstract descriptions of auditory spatial receptive fields. Basis function (A), is most strongly informative about position of the sound source behind the head (around $180$ degrees), which is also reflected in the timecourse of its activation. The Fisher information peaks in visually inaccessible areas also in other, depicted basis functions (subfigures (B), (C), (E)). There, however, the peak is not as pronounced as in the first basis function, and sensitivity to frontal positions is also visible. Fisher information of monaural basis functions (subfigures (F)-(J)) does not reveal spatial selectivity, is order of magnitude smaller and would most probably vanish in the limit of more samples.

All binaural basis functions presented on fig \ref{fig:08} are weakly temporally modulated. Temporally modulated basis functions, do not correlate strongly with the speaker's position (they also did not contribute to the position decoding, as described in the previous section).

\section{Discussion}
Experimental evidence suggests that redundancy across neurons is decreased in the consecutive processing stages in the auditory system \cite{Chechik}. This is directly in line with the efficient coding hypothesis. Additionally, a number of studies have provided evidence of adaptation of binaural neural circuits to the statistics of the auditory environment over different time scales. 
Harper and McAlpine \cite{HarperMcAlpine} have argued that properties of neural representations of interaural time differences across many species can be explained by a single principle of coding with maximal accuracy (maximization of Fisher information). Two recent experimental studies provide physiological and psychophysical evidence that neural representations of interaural level \cite{Dahmen}, and time \cite{Maier} disparities are not static but adapt rapidly to statistics of the stimulus ensemble.
The present study contributes to this lines of research, by showing that coding of the auditory space can be achieved by redundancy reduction of spectrotemporal sound representation.

The auditory system has to infer the spatial arrangement of the surrounding space by analyzing spectrotemporal patterns of binaural sound.  Auditory spatial receptive fields are formed, by extracting signal features which correlate well with environment's spatial states and result from the head related filtering.
Both sound datasets used in the present study included two, categorically different variability sources: spatial information carried by binaural differences resulting from the HRTF filtering and the raw sound waveform. Application of ICA - a simple redundancy reducing transform led to a separation of those information sources and formation of distinct model neuron sub-populations with specific spatial and spectrotemporal sensitivity.

\subsection{Linear processing of spectrotemporal binaural cues}
Emulation of the cochlear processing by performing spectral decomposition and application of the logarithmic nonlinearity produces a data representation well adapted for the position decoding task. While it is usually argued that the logarithmic nonlinearity implemented by mechanical response of the cochlear membrane is useful for reducing the dynamical range of the signal \cite{RoblesRuggero} it provides an additional advantage. Since in the frequency domain convolution is equivalent to a pointwise product of the signal and the filter \cite{Katznelson}, a logarithm transforms it to a simple addition. A linear operation on the ''cochlear'' data representation suffices to extract features imposed by the pinnae filtering \cite{HarperOlshausen}. One should note, however, that in complex listening situations involving  more than a single, stationary sound source, this simple relationship (as described by equation \ref{logsum}) may be distorted and extracted features can be mixing different aspects of the signal.

It has been observed that a linear approximation of spectrotemporal receptive fields in the auditory cortex predicts their spatial selectivity \cite{Schnupp}. This result may be surprising given that sound localization is a non-linear operation \cite{Jane} and that in a general case, linear STRF models do not explain firing patterns of auditory neurons \cite{Escabi, Cristianson}. Results shown in this paper suggest that a linear-redundancy reducing transform applied to log-spectrograms suffices to create model spatial receptive fields, providing a candidate computational mechanism explaining results provided by \cite{Schnupp}.
Localization of a natural sound source involves information included in multiple frequency channels. Binaural cues such as ILD are computed in each channel separately and have to be fused together at a later stage. This is exemplified by temporally constant basis functions learned using simulated and natural datasets. They linearly weight levels in frequency channel of both ears and in this way form their spatial selectivity. Interestingly the weighting is often asymmetric (which is reflected by BSI values different from $-1$). Such patterns represent binaural level differences coupled across multiple frequency channels. A recent study has shown that a similar computational strategy underlies spatial tuning of binaural neurons in the nucleus of the brachium of inferior colliculus (IC) in monkeys \cite{Slee}. Since it has already been suggested that IC neurons code natural sounds efficiently \cite{Carlson}, present results extend evidence in support of this hypothesis.

\subsection{Complex shapes of binaural STRFs}
Early binaural neurons localized in the auditory brainstem can be classified according to kinds of input they receive from each ear (inhibitory-excitatory - IE and excitatory-excitatory - EE) \cite{Grothe}. At the higher stages of auditory processing (Inferior Colliculus, Auditory Cortex), binaural neurons respond also to complex spectrotemporal excitation-inhibition patterns \cite{Miller, Qiu, Schnupp}.
The present paper suggests, which kinds of binaural features may be encoded and used for spatial hearing tasks by higher binaural neurons. It demonstrates that the reconstruction of natural binaural sounds requires basis functions representing various spectrotemporal patterns in each ear. The dictionary of learned binaural features is best described by a continuous binaural similarity value (in this case Pearson's correlation coefficient - BSI) and not by a classification into non-overlapping IE-EE groups. Temporally modulated basis functions consitute a particularly interesting subset of all binaural ones. Many of them represent a single cycle of envelope modulation, in opposite phase in each ear (see figure \ref{fig:06} (D)). The time interval corresponding to such phase shift is, however, much larger than the one required for the soundwave to travel between the ears. Their emergence and aspects of the environment they represent remain to be explained. 
Coding of different spectrotemporal features in each ear is useful not only for sound localization and tracking, but may be also applied for separation of sources while parsing natural auditory scenes (i.e. solving the ''cocktail party problem'').

\subsection{The role of HRTF structure}
\begin{figure}
\begin{center}
\includegraphics[width=0.5\textwidth]{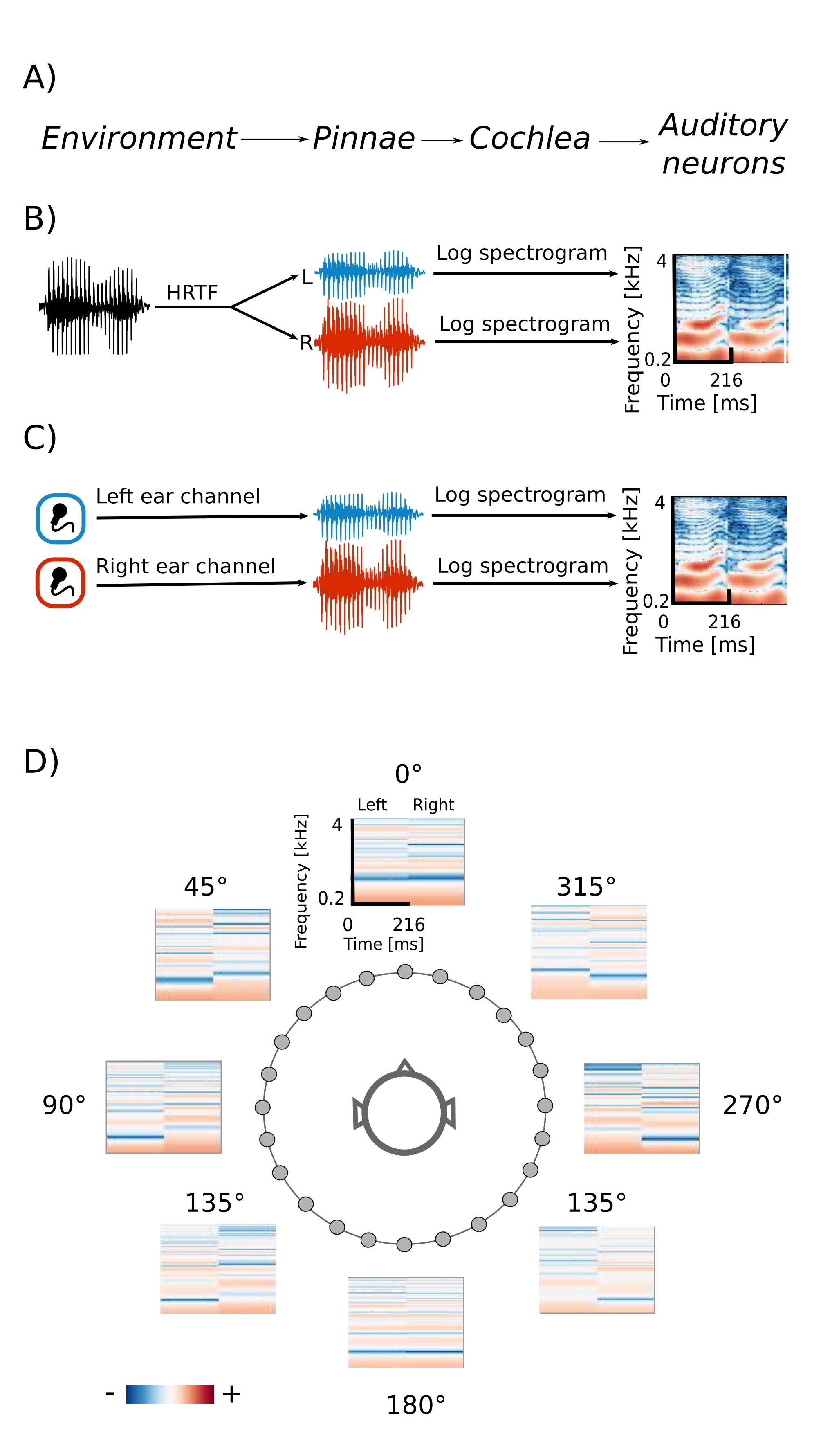}
\end{center}
 \textbf{\refstepcounter{figure}\label{fig:01} Figure \arabic{figure}.}{ Data generation process. (A) Interpretation of consecutive stages of data generation. The acoustic environment is either simulated (B) or recorded with binaural microphones (C). Further stages of the processing include frequency decomposition and transformation with a logarithmic nonlinearity, which emulates cochlear filtering (D) Positions of HRTFs around the head are marked with circles}
\end{figure}

\begin{figure}
\begin{center}
\includegraphics[height=0.9\textheight]{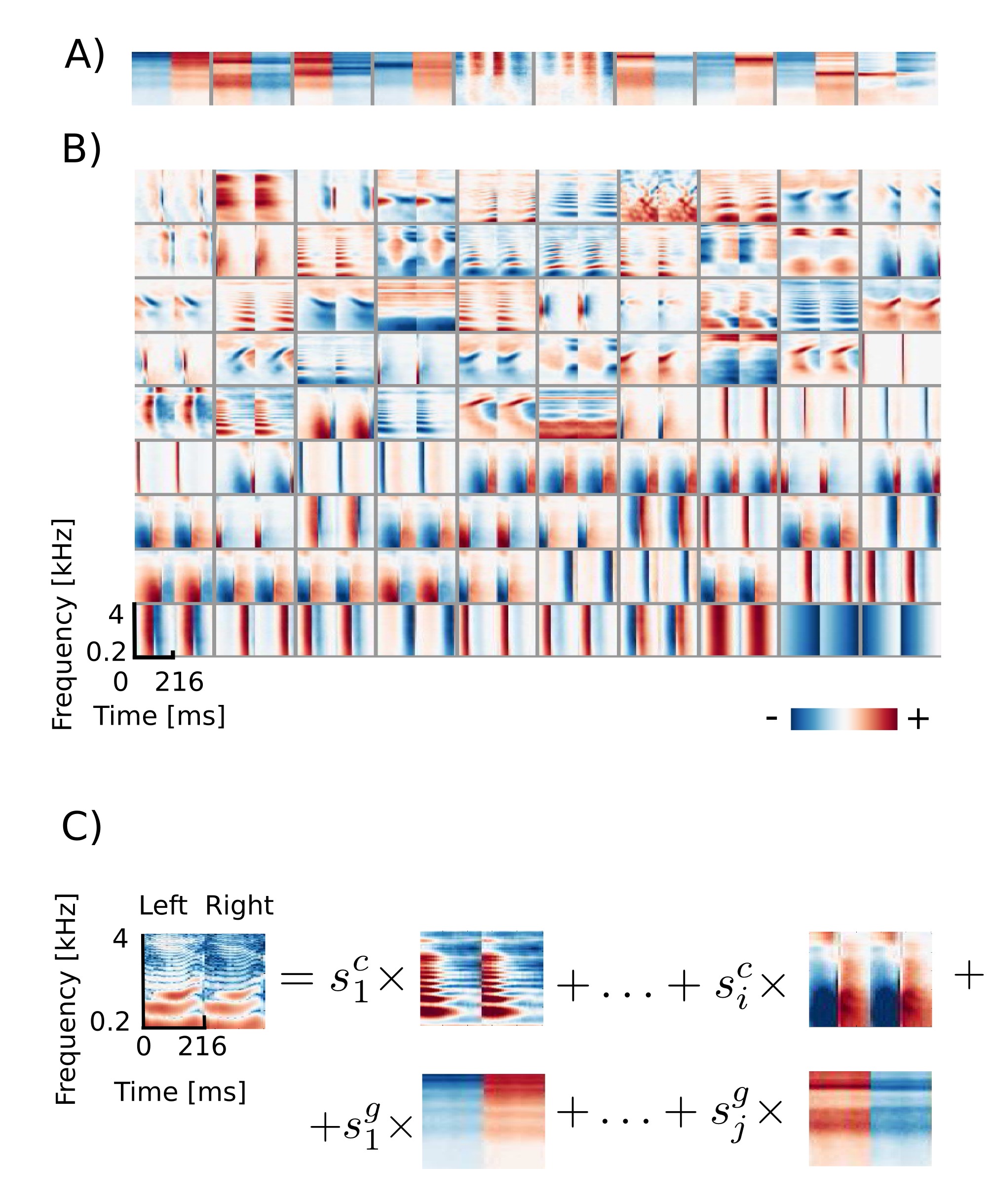}
\end{center}
 \textbf{\refstepcounter{figure}\label{fig:02} Figure \arabic{figure}.}{ ICA basis functions trained on simulated sounds (A) Binaural basis functions $a_i^g$. Left and right ear parts are dissimilar. (B) Monaural basis functions $a_i^k$. Left and right ear parts are highly similar. (C) Explanation of the representation. Each stimulus can be decomposed into a linear combination of monaural basis functions (multiplied by their coefficients $s_i^c$) and binaural ones multiplied by coefficients $s_i^g$.}
\end{figure}

\begin{figure}
\begin{center}
\includegraphics[width=\textwidth]{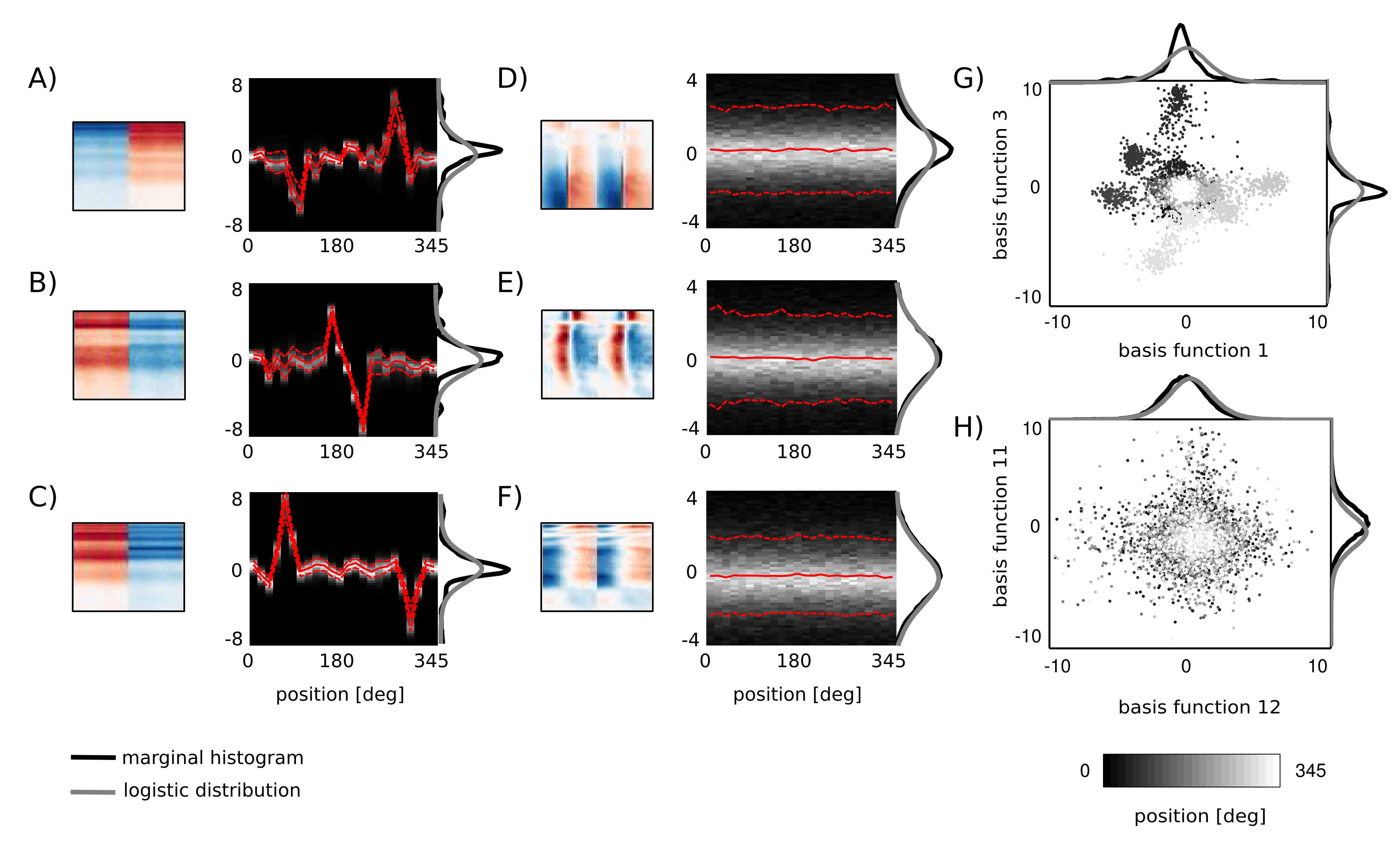}
\end{center}
 \textbf{\refstepcounter{figure}\label{fig:03} Figure \arabic{figure}.}{Spatial sensitivity of basis functions. (A)-(F) Spectrotemporal basis functions and associated conditional histograms of linear coefficients $s$. Solid red lines mark means and dashed lines limits of plus/minus standard deviation. (G)-(H) Example pairwise dependencies between monaural and binaural basis functions respectively. Each point is one sound and grayscale corresponds to its spatial position}
\end{figure}

\begin{figure}
\begin{center}
\includegraphics[width=\textwidth]{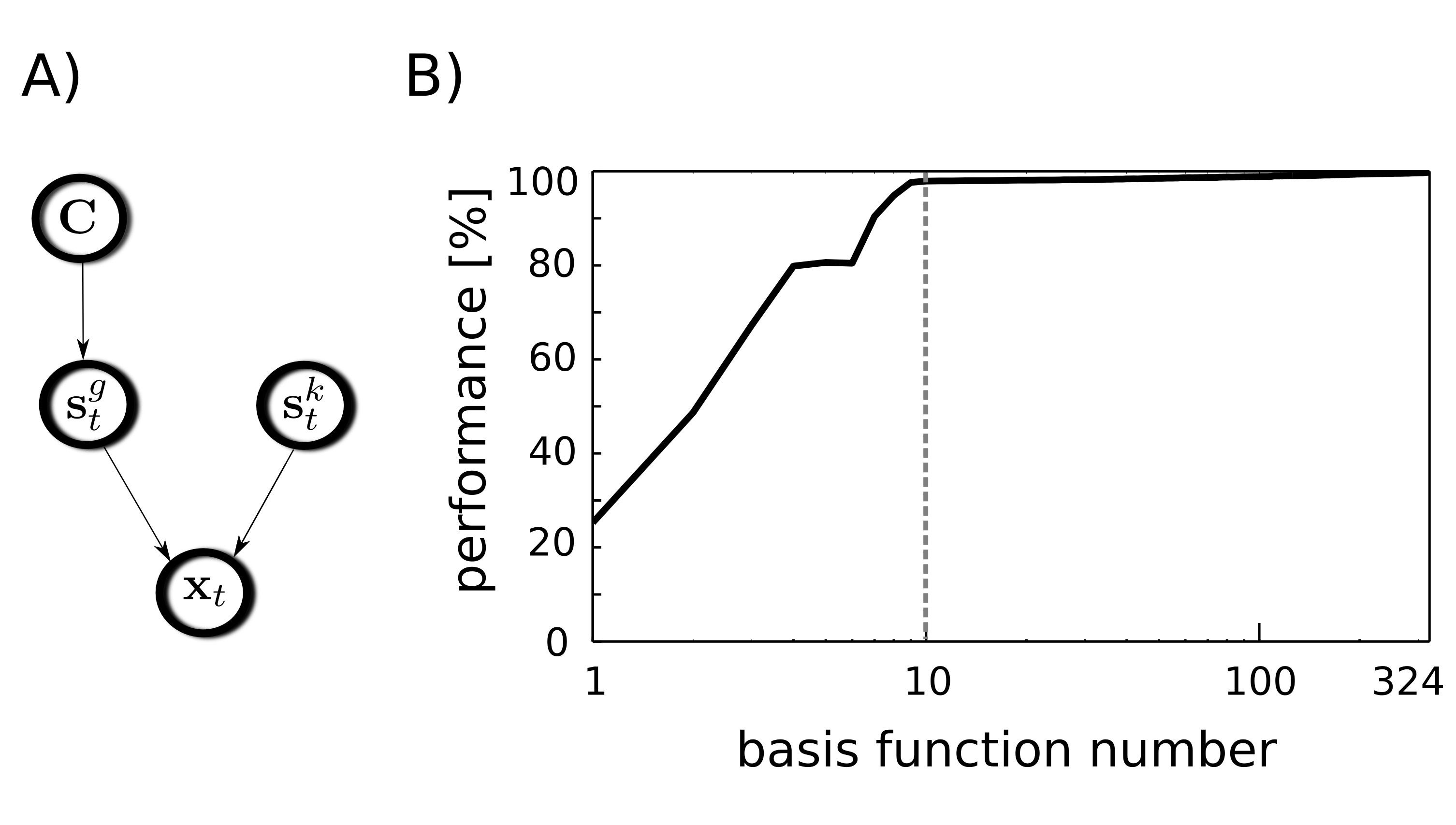}
\end{center}
 \textbf{\refstepcounter{figure}\label{fig:04} Figure \arabic{figure}.}{ Position decoding model.(A) - A graphical model representing variable dependencies. (B) - Decoders performance plotted against the number of used basis functions. Vertical dashed line separates binaural basis functions from monaural ones.}
\end{figure}

\begin{figure}
\begin{center}
\includegraphics[width=\textwidth]{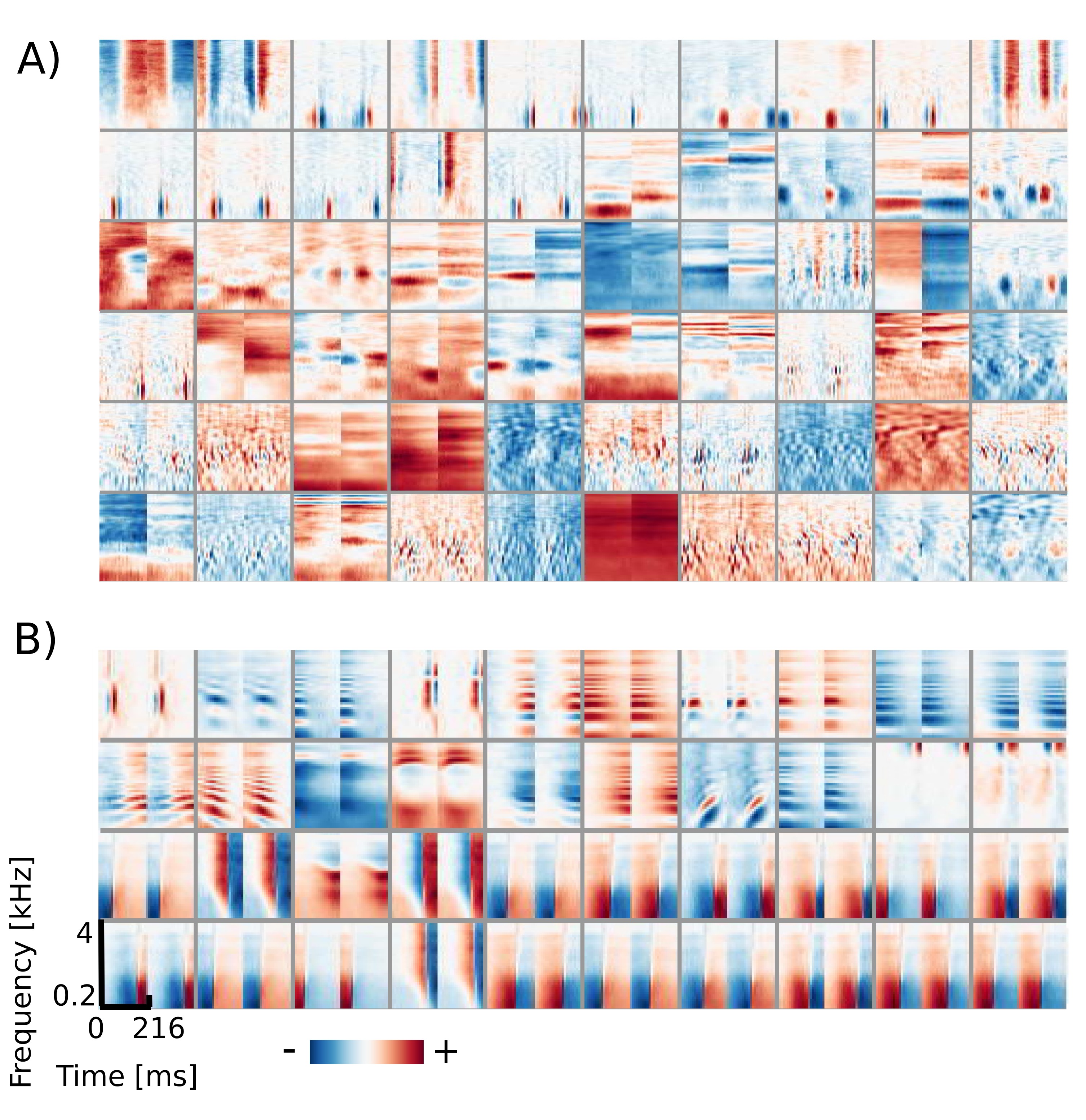}
\end{center}
 \textbf{\refstepcounter{figure}\label{fig:05}Figure \arabic{figure}.}{ Basis functions learned using natural data. (A) - Binaural basis functions ($60$ out of $64$) (B) - Monaural basis functions ($40$ out of $250$)}
\end{figure}

\begin{figure}
\begin{center}
\includegraphics[width=\textwidth]{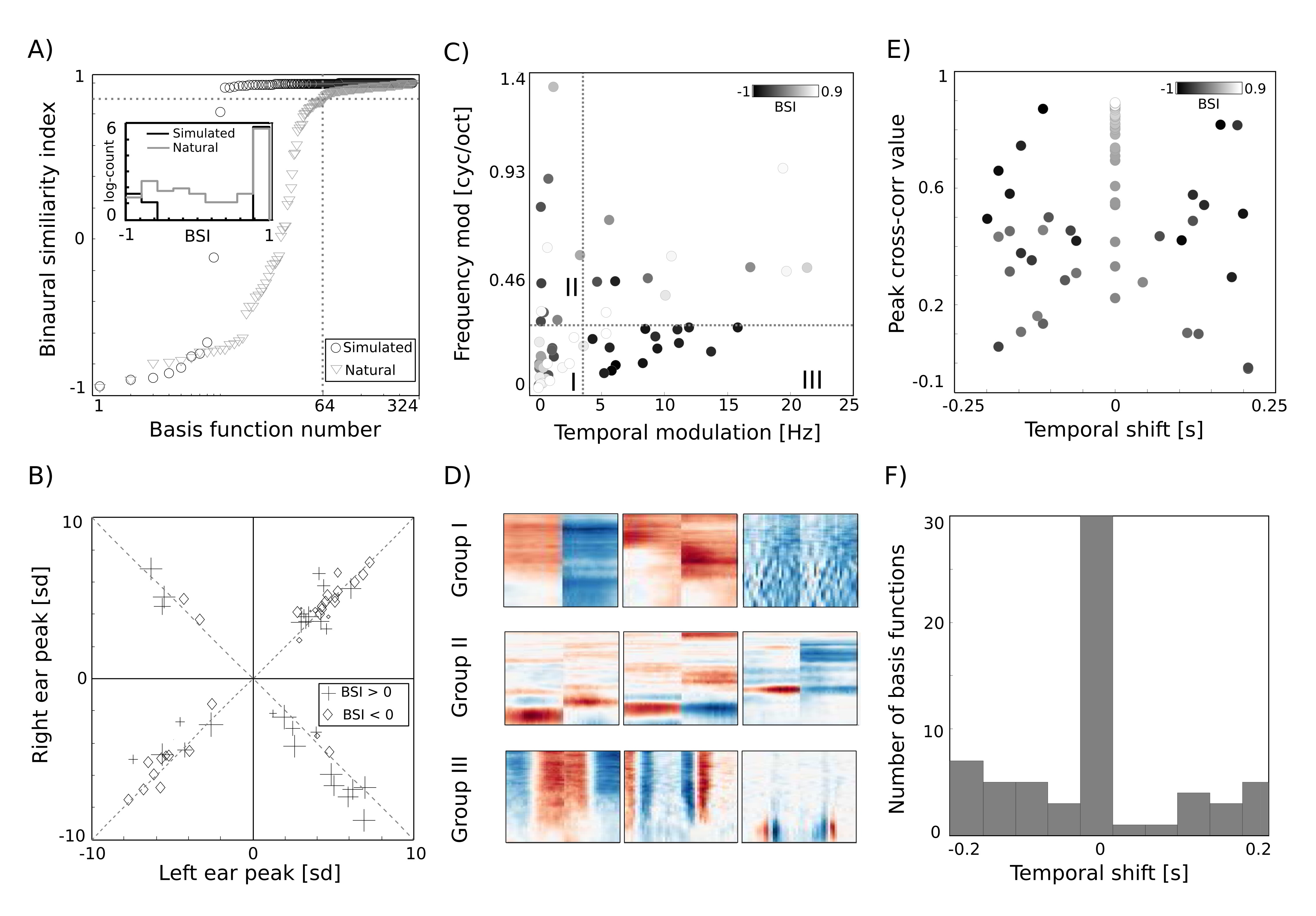}
\end{center}
 \textbf{\refstepcounter{figure}\label{fig:06} Figure \arabic{figure}.}{ Properties of basis functions learned using natural data. (A) - BSI values of natural and simulated bases. (B) - Peak values of binaural bases. (C)- Centers of mass of modulation spectra (D) - Exemplary basis functions belonging to groups I, II and III (E) - Temporal cross-correlation plotted against its peak value. Color marks the BSI (F) - A histogram of temporal shifts maximizing the cross correlation}
\end{figure}

\begin{figure}
\begin{center}
\includegraphics[width=\textwidth]{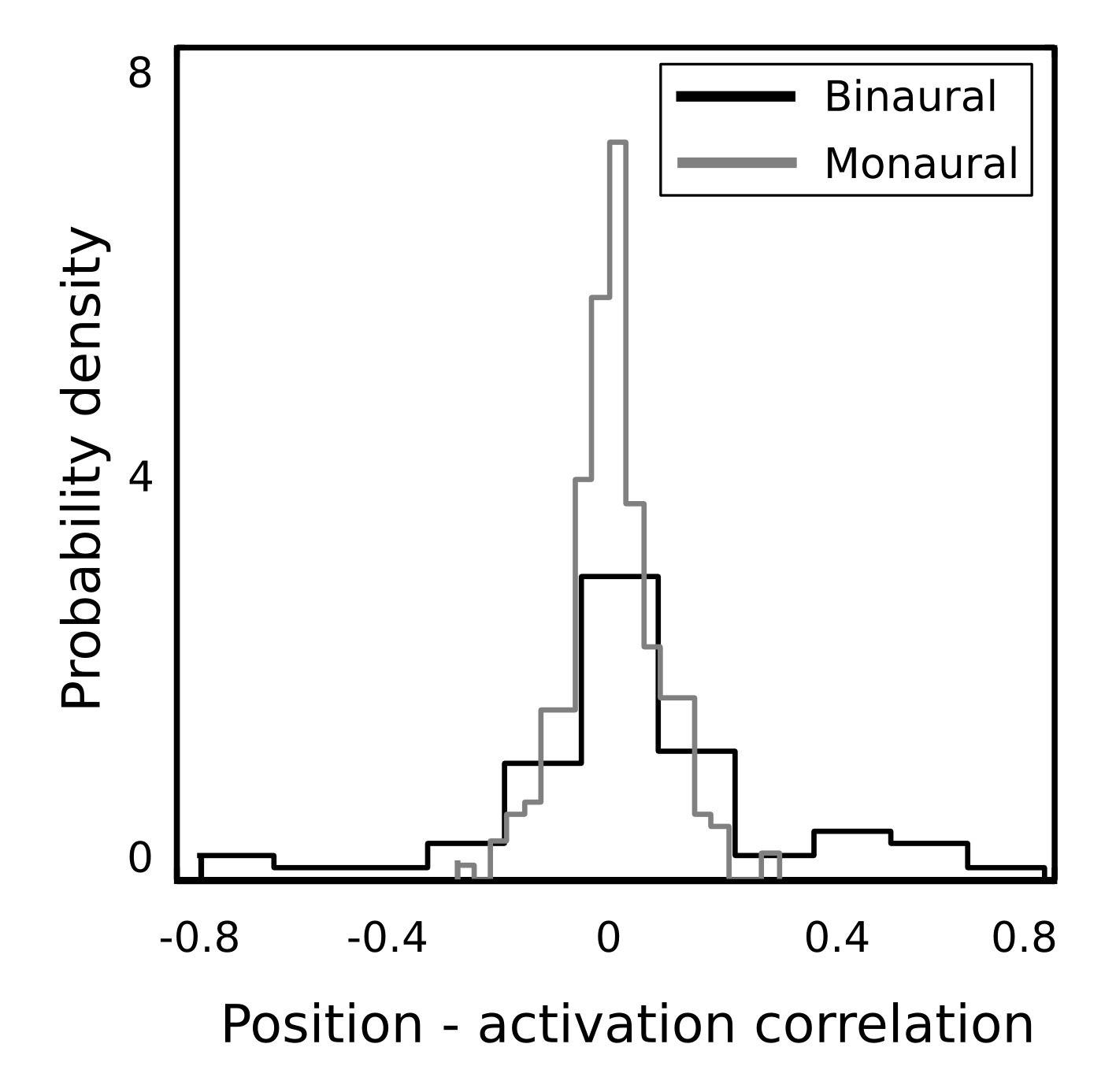}
\end{center}
 \textbf{\refstepcounter{figure}\label{fig:07}Figure \arabic{figure}.}{ Normalized histograms of activation-position correlations}
\end{figure}

\begin{figure}
\begin{center}
\includegraphics[width=\textwidth]{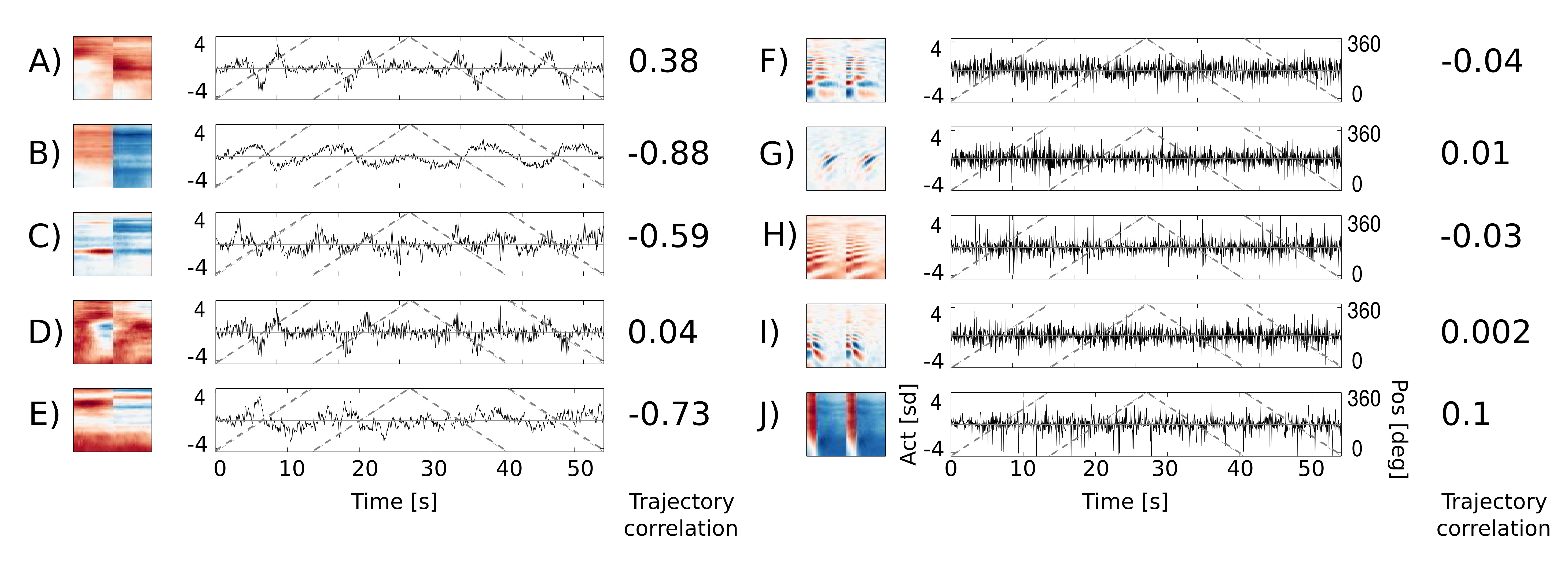}
\end{center}
 \textbf{\refstepcounter{figure}\label{fig:08}Figure \arabic{figure}.}{ Activation time course of basis functions learned using natural data.An audio-video version is available in the supplementary material. Subfigures (A)-(E) depict binaural basis functions with their activation time courses, while subfigures (F)-(J) monaural ones. Black continuous lines mark standarized activation values, gray dashed lines mark speaker's angular position.}
\end{figure}

\begin{figure}
\begin{center}
\includegraphics[width=\textwidth]{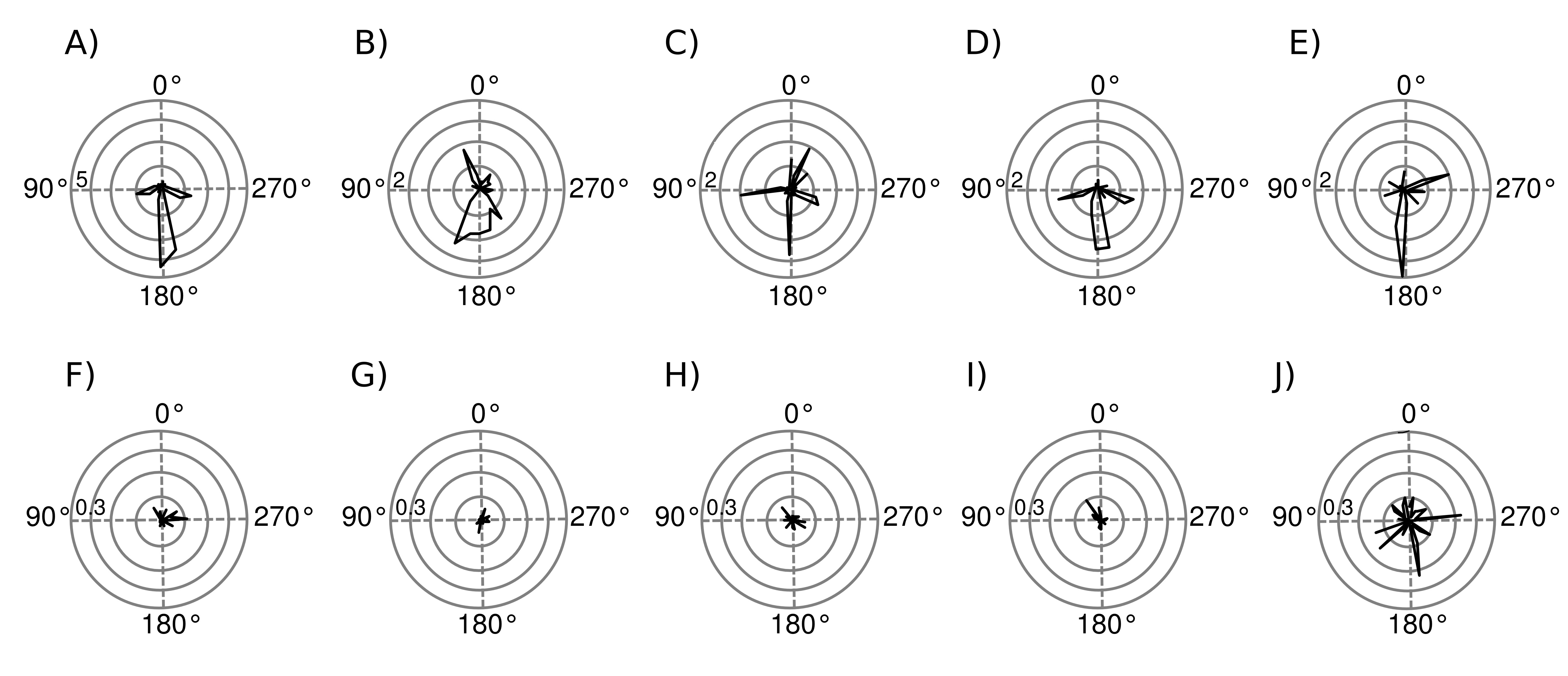}
\end{center}
 \textbf{\refstepcounter{figure}\label{fig:09} Figure \arabic{figure}.}{Spatial sensitivity quantified with Fisher information. Polar plots represent area surrounding the listener, black lines mark Fisher information $\mathcal{I}(\theta)$ at each angular position. Each subfigure corresponds to a basis function on the previous figure marked with the same letter. Please note different scales of the plots.}
\end{figure}

Spatial information is created when the sound waveform becomes convoluted with the head and pinnae filter - HRTF. By taking into account that this convolution is equivalent to addition of the log-spectral representation of the sound and the HRTF, one may conclude that the ICA recovers exact HRTF forms. A subset of basis functions learned by the ICA model from the simulated data could, in principle, contain $24$ elements, which would constitute an exactly recovered set of HRTFs used to generate the training data (see figure \ref{fig:01} D). The other basis function subset would contain features modeling speech variability. This is, however, not the case. Firstly - in the simulated dataset - HRTFs corresponding to $24$ positions were used, $10$ basis functions emerged and only $8$ were temporally non-modulated, as HRTFs are. Despite such dimensionality reduction, information included in the $8$ basis functions was sufficient to perform the position decoding with $15 \deg$ spatial resolution. This implies that binaural basis functions did not recover HRTF shapes but rather formed their compressed representation. It is important to note here that learned binaural features were much smoother and did not include all spectral detail included in HRTFs themselves (compare basis binaural basis functions from figures \ref{fig:02} A and \ref{fig:05} A with HRTFs from figure \ref{fig:01} D). The fact that coarse spectral information suffices to perform position decoding stands in accord with human psychophysical studies. It has been demonstrated that HRTFs can be significantly smoothed without influencing human performance in spatial auditory tasks \cite{Kulkarni}.

In humans and many other species, the area behind the listener's head is inaccessible to vision and information about the presence or motion of objects there can be obtained only by listening. This particular spatial information is of high survival value since it may inform about an approaching predator. Interestingly, in both used datasets features providing pronounced information about presence of sound sources behind the head clearly emerged (see figs \ref{fig:03} (B) and \ref{fig:08} (A)). Their sensitivity to sound position quantified with Fisher information is highest for the area roughly between $160$ to $230$ degrees. Since those basis functions reflect the HRTF structure, one could speculate that the outer ear shape (which determines the HRTF) was adapted to make this valubable spatial information explicit. It is interesting to think that one of the factors in pinnae evolution, was to provide spectral filters, highly informative about sound positions behind the head. This, however, can not be verified within the current setup and remains a subject of the future research.

\subsection{Conclusion}
Taken together, this paper demonstrates that a theoretical principle of efficient coding can explain the emergence of functionally separate neural populations. This is done using an exemplary task of binaural hearing. 

In a previous work by \cite{Asari}, it has already been shown that a sparse coding approach may be useful for monaural position decoding. Their work, however, did not show separation of spatial and identity information into two distinct channels. Additionally in contrast to this study, \cite{Asari} relied on simulated data which did not include patterns such as sound motion.

Learning independent components of binaural spectrograms has extracted spatially informative features. Their sensitivity to sound source position, was therefore a result of applying a general-purpose strategy. This may suggest that seemingly different neural computations may be instantiations of the same principle. For instance, it may appear that auditory neurons with spatial and spectrotemporal receptive fields must perform different computations. Current results imply that they may be sharing a common underlying design principle - efficient coding, which extracts stimulus features useful for performing probabilistic inferences about the environment. 

\section*{Disclosure/Conflict-of-Interest Statement}
The author declares that the research was conducted in the absence of any commercial or financial relationships that could be construed as a potential conflict of interest.

\section*{Acknowledgement}
I would like to thank Nils Bertschinger, Timm Lochmann, Philipp Benner and Pierre Baudot for helpful discussions and proofreading of this paper. 

\paragraph{Funding} 
This work was funded by DFG Graduate College ''InterNeuro''.

\section*{Supplemental Data}
Movie1.avi - a multimedia version of figure 8

\bibliographystyle{plain}
\bibliography{test}

\section*{Figures}

\end{document}